# Improving Students' Understanding of Magnetism

Chandralekha Singh
*Department of Physics and Astronomy, University of Pittsburgh, Pittsburgh, PA 15260*

**Abstract**

Formally investigating the sources of students' difficulties around specific subjects is crucial for developing appropriate strategies to help students. We have been studying difficulties in understanding magnetism encountered by students in the calculus-based introductory physics courses. A majority of these students are engineering majors. Student difficulties were assessed by administering written free-response questions and by developing and administering a research-based multiple-choice test to students. We also conducted individual interviews with a subset of students. Some of these interviews were lecture-demonstration based interviews in which students were asked to predict the outcomes of experiments, perform the experiments and reconcile the differences between their predictions and observations. Some of the common misconceptions found in magnetism are analogous to those found in electrostatics. Some additional difficulties are due to the non-intuitive three dimensional nature of the relation between magnetic field, magnetic force and velocity of the charged particles or direction of current. Another finding is that students often used their gut feeling and had more difficulty reconciling with the idea that the magnetic force and field are perpendicular to each other when they were shown actual lecture-demonstration setups and asked to predict outcomes of experiments than when they were asked to explain relation between magnetic force and field theoretically solely based upon an equation. We discuss the implications of this research to teaching and learning.

**Introduction**

Magnetism is an important topic covered in most calculus-based introductory physics courses for science and engineering majors. The Physics Education Research literature is full of studies of student difficulties in introductory mechanics.[1] Student difficulties in Electricity and Magnetism have not received nearly as much attention[2-8]. The Washington group under the direction of McDermott has developed tutorials on electric field, potential, potential energy, and magnetism related mostly to bar magnets[3]. Maloney et al.[9] and Ding et al.[10] have developed broad surveys to evaluate students' conceptual knowledge of all of electricity and magnetism covered in introductory physics.

Here, we discuss research on students' difficulties with concepts related to magnetism covered in introductory calculus-based physics courses. The origins of student difficulties in learning physics concepts can broadly be classified into two categories: gaps in students' knowledge, and misconceptions. Cognitive theory suggests that learning is incremental and new knowledge builds on prior knowledge[11, 12]. Knowledge gaps can arise from many sources, e.g., a mismatch between the level at which the material is presented in a course and students' prior knowledge.

Deep-rooted misconceptions can also seriously impede the learning process at all levels of instruction[1]. Misconceptions can sometimes arise from naive generalizations or over-generalizations of concepts learned in one context to another, or from a compounding of misconceptions that were never cleared up at an earlier stage. In introductory courses, misconceptions can arise sometimes due to the fact that students try to rationalize their everyday experiences in terms of their limited knowledge base, and because many of their justifications and rationalizations do not conform to the laws of physics.

Our study was conducted using two methods:

- design and administration of diagnostic tools, e.g., pre/post free-response and multiple-choice tests
- in-depth interviews with individual students using a think-aloud protocol[13].

The major advantage of written tests was that we could administer them to students in many large introductory physics courses. We preceded the development of the research-based multiple-choice test with written free-response questions requiring students to explain their reasoning explicitly. We began by administering carefully designed free response questions to a large number of students. Based upon the answers to the free response questions, the incorrect alternative choices to the multiple-choice questions were designed incorporating the most common difficulties students had about those concepts. Multiple-choice tests were useful for collecting data from a large number of students because such tests are easy to grade and analyze quantitatively. Although the thought process is not revealed completely by the multiple-choice test answers alone, when students are asked for explanations after each question and when the test is administered in conjunction with in-depth interviews of a subset of students, such tests can be very helpful in pinpointing student difficulties. We conducted hour-long individual interviews with 25 students using the think-aloud protocol. We allowed students to express their views without interruption after asking them questions during an interview. Afterwards, we probed them for clarification of points not explained clearly in their initial responses. Fifteen of these individual interviews were geared towards asking students to answer free-response or multiple-choice questions during the development of the multiple-choice tests. The other ten interviews involved asking students to answer part of the written questions similar to those given to the other fifteen students but also asking them to predict the outcomes of various experiments related to the questions as appropriate[13]. After the initial predictions of what should happen, students performed the experiment and were asked to reconcile the differences between their initial responses and the observation. Since some students participated in lecture-demonstration based interviews while others only did interviews with paper-pencil tasks, we had an opportunity to compare these two groups.

**Results and Discussion**

The details about the development of the 30 item conceptual magnetism diagnostics test, which takes 50 minutes to administer, will be discussed elsewhere. In some classes, students were asked to explain their reasoning for each question. The magnetism topics covered in the test included magnitude and direction of the magnetic field produced by current carrying wires, forces on current carrying wires in an external magnetic field, force and trajectory of a charged particle in an external magnetic field, work done by the external magnetic field on a charged particle, magnetic field produced by bar magnets, force between bar magnets and static charges. Below, we will only discuss students' responses to selected multiple-choice questions before (pre-test) and after (post-test) college introductory physics instruction in relevant concepts. The pre-test was given in the first week of recitation class and students were told explicitly that the goal of the pre-test was to evaluate their prior knowledge about magnetism from high school or everyday experience. They were told to try their best but were given assurance that if they did not know something it was fine because the concepts would be discussed during the course. The average score on the last version of the magnetism test (including all 30 questions) from calculus-based introductory physics was 28% on the pre-test (administered to two sections of the course with 127 students total) and 50% on the post-test (administered to six sections with 466 students total). In all of the courses, students were given these questions as part of a recitation quiz and it counted for a small part of their course grade. Although student performance on the post-test is better overall than their performance on the pre-test, students struggled with the concepts even after instruction. The reliability coefficient, alpha, which is a measure of the internal consistency of the test, is greater than 0.8 on the post-test which is considered good by the standards of test design[14]. The point biserial discrimination is related to the ability of a question to discriminate between students who overall did well on the test vs. those whose overall performance on the test was poor. This discrimination index for 26 items on the test was greater than 0.3, which is also considered good by the standards of test design[14]. Table 1 shows the average student performance before and after instruction on selected questions from the test discussed below. These questions address students' difficulties in distinguishing between an electric dipole and magnetic dipole, forces between current carrying wires and the magnetic field produced by them and the force on a charged particle in an external magnetic field. The correct choice for each multiple-choice question is italicized. As we discuss the difficulties and misconceptions, we will also briefly describe relevant findings from individual interviews including those involving lecture-demonstration based interviews.

| Question | 1 | 2 | 3 | 4 | 5 | 6 | 7 |
|---|---|---|---|---|---|---|---|
| Pre-test | 38 | 29 | 39 | 17 | 42 | 24 | 16 |
| Post-test | 41 | 55 | 39 | 64 | 47 | 43 | 49 |

TABLE I: The percentage of correct responses on the selected questions discussed here on the pre-test (before instruction in calculus-based introductory physics course in two sections with 127 students) and post-test (466 students from 6 sections of the course).

### Distinguishing between Electric Dipole and Magnetic Dipole

The first question below exemplifies students' difficulties distinguishing between an electric dipole and a magnetic dipole:

Question (1): An electric dipole and a bar magnet with the same length $L$ are both at rest. They are equidistant from the y-axis, and each is symmetrically placed about the x-axis (see Figure).

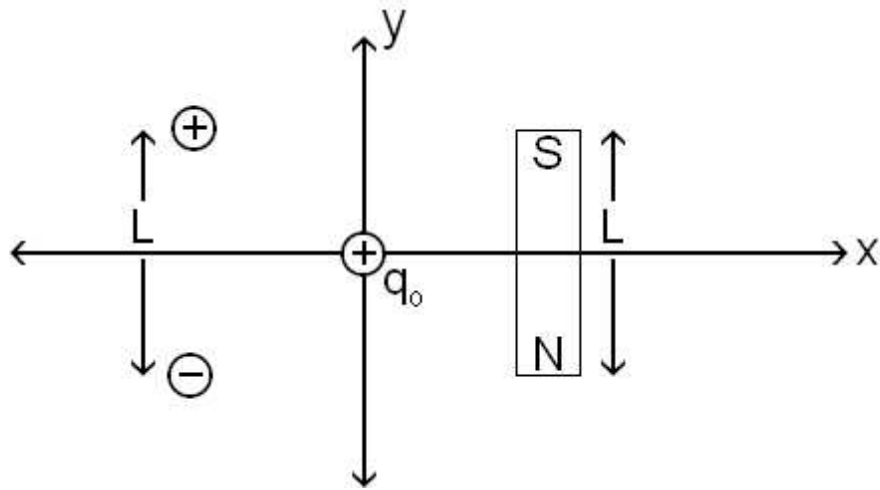

A positive point charge $q_0$ is at rest at the origin as shown. In what direction is the net force on the charge $q_0$ due to the electric dipole and the bar magnet?

(a) positive y direction

(b) *negative y direction*

(c) positive x direction

(d) negative x direction

(e) There is no net force on $q_0$ due to the electric dipole and the bar magnet.

As can be seen from Table 1, the post-test performance on question (1) is 41% which is not significantly different from the pre-test performance of 38% (before instruction). The most common incorrect response was option (e), i.e., there is no net force on charge $q_0$. Both written explanations and interviews suggest that students often have difficulty distinguishing between the poles of a magnet and point charges. As a result, many students claimed that a bar magnet is an electric dipole with positive and negative charges located at the north and south poles. Thus, they incorrectly concluded that a point charge at rest close to a static bar magnet will feel a force due to the magnet. When the interviewer asked them whether a magnet will retain its magnetic properties when broken into two pieces by cutting it at the center, most of them claimed that the magnetic properties will be retained, and each of the two smaller pieces will become a magnet with opposite poles at the ends. When the interviewer asked how that is possible if they had earlier claimed that the north and south poles are essentially localized opposite charges, some students admitted they were unsure about how to explain the development of opposite poles at each end of the smaller magnets but others provided creative explanations about how charges will move around while the magnet is being cut into two pieces to ensure that each of the cut pieces has a north and south pole. During the interviews, we probed how students developed the misconception that a bar magnet must have opposite charges localized at the two poles because when a child plays with a bar magnet, he/she does not necessarily think about localized electric charges at the end. Some students claimed that they learned it from an adult while he/she tried to explain why a bar magnet behaves the way it does or they had heard it on a television program. More research is needed to understand how and at what age college students developed such incorrect notions.

**Forces between Current Carrying Wires and Magnetic Field due to them**

The following two questions deal with students' difficulties concerning the forces between current carrying wires and the magnetic field produced by current carrying wires:

Question (2): Two very long straight parallel wires (labeled 1 and 2 in the figure) are separated by a small distance and carry the same current *i*. The current in wire 1 flows out of the page, and that in wire 2 flows into the page.

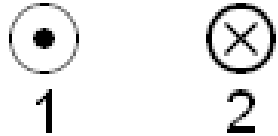

What is the direction of the force <u>on wire 1</u> due to wire 2?

   (a) Into the page ($\otimes$)

   (b) Out of the page ($\odot$)

   (c) *To the left* ($\longleftarrow$)

   (d) To the right ($\longrightarrow$)

   (e) Not enough information.

The most common incorrect response for question (2) about the direction of the force on a current carrying wire due to a parallel wire carrying current in the opposite direction was based on the assumption that the wires would attract each other (option (d)). Even in the individual lecture-demonstration based interviews, many students explicitly predicted that the wires would attract and come closer to each other. When asked to explain their reasoning, students often cited the maxim "opposites attract" and frequently made an explicit analogy between two opposite charges attracting each other and two wires carrying current in opposite directions attracting each other. In lecture-demonstration based interviews, when students performed the experiment and observed the repulsion, none of them could reconcile the differences between their initial prediction and observation. One difficulty with Question (2) is that there are several distinct steps (as opposed to simply one step) involved in reasoning correctly to arrive at the correct response. In particular, understanding the repulsion between two wires carrying current in opposite directions requires comprehension of the following issues: (I) There is a magnetic field produced by each wire at the location of the other wire. (II) The direction of the magnetic field produced by each current carrying wire is given by a right hand rule. (III) The magnetic field produced by one wire will act as an external magnetic field for the other wire and will lead to a force on the other wire whose direction is given by a right hand rule. None of the interviewed students were able to explain their reasoning systematically. What is equally interesting is the fact that none of the interviewed students who correctly predicted that the wires carrying current in opposite directions would repel could explain this observation based upon the force on a current carrying wire in a magnetic field even when explicitly asked to do so. A majority of them appeared to have memorized that wires carrying current in opposite directions repel. They had difficulty applying both right hand rules systematically to explain their reasoning: the one about the magnetic field produced by wire 2 at the location of wire 1 and the other one about the force on wire 1 due to the magnetic field. This difficulty in explaining their prediction points to the importance of asking students to explain their reasoning.

Question (3): Two very long straight parallel wires (labeled 1 and 2 in the figure) are separated by a small distance and carry the same current $i$. The current in wire 1 flows out of the page, and that in wire 2 flows into the page. What is the direction of the magnetic field at point P due to the current in the wires? Point P is equidistant from the wires as shown in the figure below.

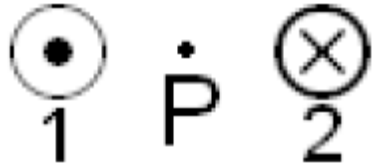

(a) To the right ($\longrightarrow$)

(b) To the left ($\longleftarrow$)

(c) *Upward* ($\uparrow$)

(d) Downward ($\downarrow$)

(e) The magnetic field is zero at point P.

As can be seen from Table 1, there was no improvement on post-test compared to pre-test on question (3). The most common incorrect response for this question was (e). These students claimed that the magnetic field is zero at point P between the wires. In written explanations and individual interviews, students who claimed that the magnetic field is zero at point P between the wires carrying opposite currents argued that at point P, the two wires will produce equal magnitude magnetic fields pointing in opposite directions, which will cancel out. During interviews, when students were explicitly asked to explain how to find the direction of the magnetic field at point P, they had difficulty figuring out the direction of the magnetic field produced by a current carrying wire using the right hand rule and in using the superposition principle to conclude correctly that the magnetic field at point P is upward. Interviews suggest that for many students, confusion about similar concepts in electrostatics were never cleared and had propagated to difficulties related to magnetism. For example, several students drew explicit analogy with the electric field between two charges of equal magnitude but opposite sign, incorrectly claiming that the electric field at the midpoint should be zero because the contributions to the electric field due to the two charges cancel out similar to the magnetic field canceling out in question (3). Some interviewed students who were asked to justify their claim that the electric field at midpoint between two equal magnitude charges with opposite sign is zero were reluctant claiming the result was obvious. Then, the interviewer told them that saying that the influence of equal and opposite charges at midpoint between them must cancel out is not a good explanation and they must explicitly show the direction of electric field due to each charge and then find the net field. This process turned out to be impossible for some of them but others who drew the electric field due to each charge in the same direction at midpoint were surprised. One of them who realized that the electric field cannot be zero at midpoint between the two equal and opposite charges smiled and said that this fact is so amazing that he will think carefully later on about why the field did not cancel out at midpoint. There is a question similar to question (3) on the broad survey of electricity and magnetism called CSEM survey.[9] However, students performed much worse on question (3) than on the corresponding question on CSEM (39% on post-test for question (3) vs. 63% for the corresponding question on CSEM[9]). One major difference is that we asked students for the direction of magnetic field at the midpoint between current carrying wires carrying current in opposite directions and many students had the misconception that the magnetic field is zero at the midpoint. Thus, they gravitated to option (e). In the CSEM survey[9], all incorrect choices were quite popular because the misconception specifically targeted by question (3) about the magnetic field being zero at the midpoint between the wires was not targeted there.

**Force on a Charged Particle in a Magnetic Field**

The following four questions illustrate the kinds of difficulties students have with questions related to the force on a charged particle in an external magnetic field:

Question (4): A particle with charge $+q$ and speed $v$ enters a region with uniform magnetic field of magnitude $B$ pointed opposite to the direction of the particle's initial velocity as shown below.

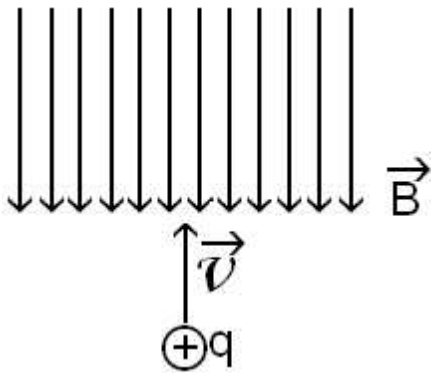

How will the motion of the particle be affected as it moves through this region?

(a) *The velocity of the particle will remain unchanged.*

(b) The particle will speed up.

(c) The particle will slow down.

(d) The particle will be deflected to the left (⟵).

(e) The particle will be deflected to the right (⟶).

In response to the question (4) above, students performed significantly better on the post-test compared to pre-test. The most common incorrect response to question (4) was option (c) followed by options (d) and (e). Interviewed students were asked such questions in the context of a lecture-demonstration related to the effect of bringing a powerful bar magnet from different angles towards an electron beam (including the case where the magnetic field and velocity vectors are collinear as in question (4)). Students were explicitly told to predict the outcome for both the electron beam and a beam of positive charges. Students who incorrectly chose option (c) in question (4) explained that the particle will slow down because the magnitude of the magnetic field is opposite to the direction of velocity and this implies that the force on the charged particle must be opposite to the velocity. Of course, this prediction could not be verified by performing the experiment. Students who chose options (d) or (e) often incorrectly remembered the right hand rule about magnetic force being perpendicular to magnetic field and provided animated explanations such as "Oh, the electron wants to get out of the way of the magnetic field and that's why it bends". They were surprised to observe that the deflection of the electron beam is negligible when the strong magnet has its north pole or south pole pointing straight at the beam.

Question (5): Consider three identical positively charged particles, labeled 1, 2 and 3. They all move with the same speed in the plane of the paper but in different directions as shown. The particles enter a region with uniform magnetic field of magnitude $B$ that is pointing into the paper.

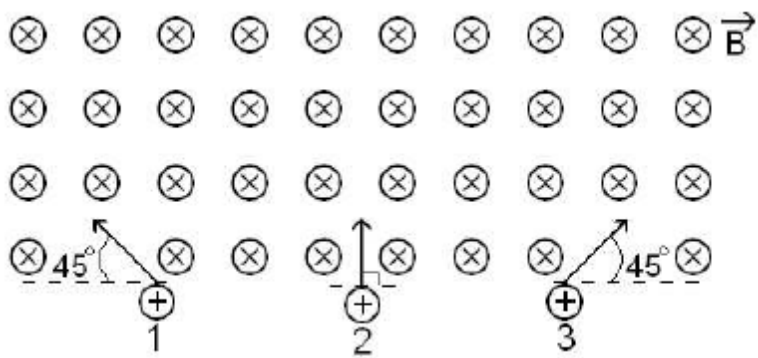

The magnetic field will exert the greatest force on which particle(s)?

(a) 1

(b) 2

(c) 3

(d) 1 and 3

(e) *The magnitude of the forces on all three particles is the same.*

Table 1 shows that there is no significant difference in the average performance on the pre-test and post-test on question (5). The most common incorrect choice for question (5) was option (b) because students used the redundant information about angle provided and had difficulty visualizing the problem in three dimensions. The correct answer is option (e) because the velocity of all of the three charged particles is perpendicular to the magnetic field. Written explanations and interviews suggest that some students incorrectly used the superfluous information provided about the angles that the charged particles (1) and (3) make with the horizontal. During interviews, only when the students choosing option (b) were asked explicit questions about the direction of the magnetic field, velocity and the angle between them for particles (1) and (3), did they realize that the angle between them is $90^0$ in all three cases. The fact that these students initially claimed that the angle between the magnetic field and velocity vector is $45^0$ for particles (1) and (3) is similar to our finding in mechanics. In the context of kinematics, when introductory physics students were asked a question about a rabbit's motion whose ears were at $45^0$ angle to the horizontal, approximately one third of the students used this redundant information about the angle of rabbit's ears to solve the problem incorrectly.

Question (6): The figure below shows two situations in which a positively charged particle moving with speed $v$ through a uniform magnetic field of magnitude $B$ experiences a magnetic force $\vec{F}_B$ due to the field.

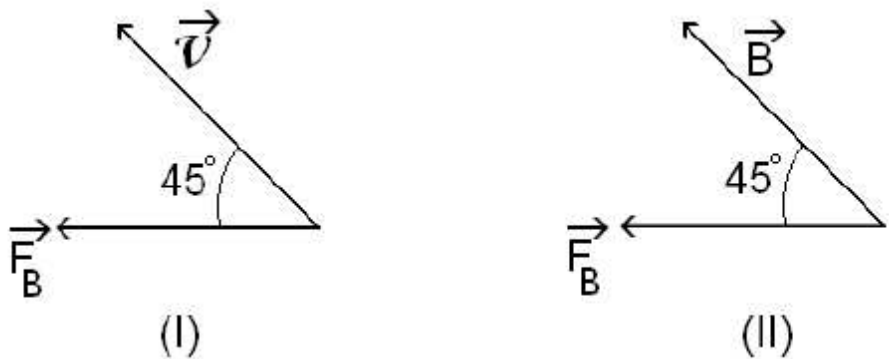

Which orientations are physically reasonable?

(a) I

(b) II

(c) Both I and II.

(d) *Neither of them.*

The student performance on this question changed from 24% on pre-test to 43% on post-test. The most common incorrect response to this question was option (a) (but options (b) and (c) were also chosen). In written explanations, students who chose option (a) often incorrectly claimed that the velocity of a charged particle and the magnetic force can be at any angle to each other, but some claimed that the magnetic field and force must be perpendicular while others said that the field and force must be parallel. In interviews, students who provided incorrect responses were asked to write down an expression for the magnetic force on the charged particle in an external magnetic field. Approximately half of the students were unable to write the correct expression. Those who wrote the correct expression were explicitly asked about the cross product between the velocity and magnetic field and what it implies about the angle between the magnetic force and the velocity or the magnetic field vector. Approximately half of these students corrected their initial error and said that the cross product implies that the magnetic force must be perpendicular to both the velocity and the magnetic field. Others did not know that the cross product of two vectors must be perpendicular to each of the vectors.

Question (7): A positively charged particle with a speed $v$ is in a region with non-zero electric and magnetic fields (with magnitudes $E$ and $B$ respectively) pointing in opposite directions as shown.

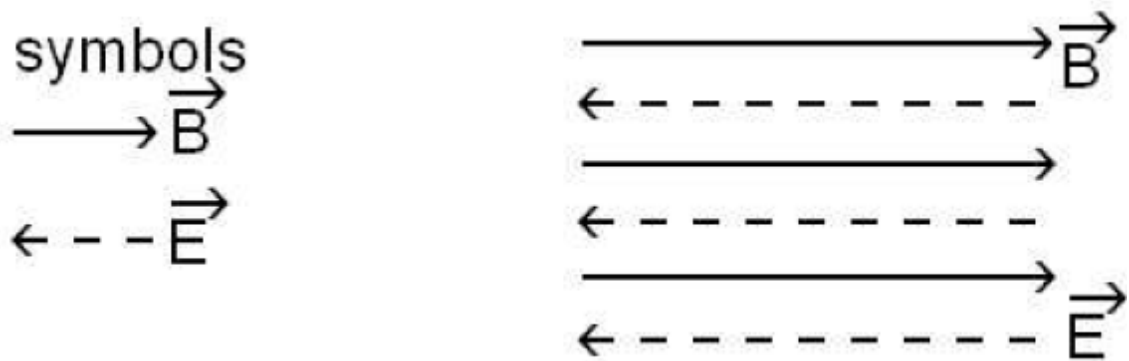

Which one of the following statements is true about the net force on the particle due to the fields shown?

(a) The net force is zero if the particle's velocity is the same as the direction of $\vec{E}$.

(b) The net force is zero if the particle's velocity is the same as the direction of $\vec{B}$.

(c) The net force is zero if the particle's velocity is perpendicular to both the $\vec{E}$ and $\vec{B}$ fields.

(d) The net force is zero if the particle's velocity is zero.

(e) *The particle always feels a net force.*

Table 1 shows that 49% of students provided the correct response on post-test, three times the number on the pre-test. The most common incorrect option in response to question (7) above was option (c). Written explanations and interviews suggest that students had learned about velocity selectors in which the electric and magnetic fields are perpendicular to each other (not the situation shown here) and to the velocity of the particles and there is a particular speed $v=E/B$ for which the net force on the particle is zero. These students had over-generalized this situation and had simply memorized that whenever the velocity is perpendicular to both the electric and magnetic fields, the net force on the charged particle is zero. They neglected to account for the fact that the magnitudes of electric and magnetic fields must satisfy $v=E/B$ and the two fields must be perpendicular to each other and oriented appropriately for the net force on the particle to be zero. During interviews, students were often not systematic in their approach and talked about the net effect of the electric and magnetic fields simultaneously rather than drawing a free body diagram and considering the contributions of each field separately first. The systematic approach to analyzing this problem involves considering the direction and magnitude of the electric force $\vec{F}_E = q\vec{E}$ and magnetic force $\vec{F}_B = q\vec{v} \times \vec{B}$ individually and then taking their vector sum to find the net force. Some interviewed students who made guesses based upon their recollection of the velocity selector example discussed in the class claimed that the net force on the particle is zero for this situation. They were asked by the interviewer to draw a free body diagram for the case in which the charged particle is launched perpendicular to both fields. Some of them who knew the right hand rule for the magnetic force and the fact that the electric field and force are collinear ($\vec{F}_E = q\vec{E}$) were able to draw correct diagrams showing that the electric force and magnetic force are not even collinear. One of these students exclaimed: "I don't know what I was thinking when I said that the net force is zero in this case. These two are (pointing at the electric and magnetic forces in the free body diagram he drew) perpendicular and can never cancel out". Such discussions with students suggest the need to teach them a systematic approach to solving problems so that they don't treat a conceptual problem as a guessing task.

**Cognitive Conflict While Predicting Outcome of Experiments**

One general finding based upon the comparison of students who had lecture-demonstration based interviews vs. paper-pencil interviews is that non-intuitive facts, e.g., the magnetic force and field are perpendicular to each other, were difficult for students to translate from theory to real situations. The concrete experimental setup involving electron beam (moving electrons) and bar magnets prompted some students to use their gut feeling and slip back to the idea that force and field must be in the same direction rather than helping them reason during the lecture-demonstration based interviews in which they were asked to predict the direction of the magnetic force before performing the experiment. For example, students were asked to predict what happens to an electron beam if the north pole of a magnet is brought straight towards the beam. They were asked to assume that the velocity of the electron beam and the magnetic field are approximately collinear. Some students said that the beam will slow down due to the retarding force of the magnetic field pointing opposite to the velocity. The same students were also shown a setup in which a long straight wire was between the poles of a horse-shoe magnet and they were asked to predict the direction of the force on the wire when a current passed through it (they were told to assume that the magnetic field was approximately from the north to the south pole). They again predicted that the wire will get deflected towards the north or south pole (not perpendicular to the straight line joining the two poles). However, when asked to write an expression for the force on the charged particle, some of them wrote an expression involving cross product and said that the magnetic force is supposed to be perpendicular to magnetic field. After writing the expression, some appeared confused about what they had predicted and said that the reason they predicted that the wire should deflect towards one of the poles is because it does not make sense for it to deflect perpendicular to it. The deflection of the wire perpendicular to the field when the current was passed through it surprised them. One student said he would never have thought that this is what the magnetic force being a cross product involving magnetic field means. In interviews involving paper-pencil task, students who interpreted the right hand rule correctly said that the magnetic force should be perpendicular to the magnetic field. It appears that unless students are given opportunity with real setups, they may have difficulty understanding what theoretical ideas such as force and field being perpendicular may mean in real contexts. It is also interesting to note that some of these lecture-demonstrations were actually shown by instructors during the lecture but interviewed students could not clearly remember what they had observed in the class.

**Summary**

We studied the conceptual difficulties and misconceptions that students in the calculus-based introductory physics courses have related to magnetism by developing and administering a research-based multiple-choice test and by interviewing a subset of students individually. Some of the misconceptions found originate from over-generalization of correct or incorrect ideas from electrostatics while others were due to the more intricate nature of magnetic forces and fields including the need to visualize the forces and fields in three dimensions. Many interviewed students went back and forth between electrostatics and magnetism and often used incorrect analogies between static charges and current carrying wires. Written explanations also corroborate these difficulties. We find that students often have difficulty in using the right hand rule to figure out the forces on moving charges or current carrying wires in external magnetic fields and also find it difficult to figure out the magnetic fields produced by current carrying wires using the right hand rule. They often mixed up the right hand rules for figuring out the magnetic forces and fields.   Another finding is that asking students to predict what should happen in a particular situation with a concrete experimental setup confused some students rather than helping them in reasoning and prompted them to use their gut feeling. Comparison between students who interviewed with lecture demonstration setups and those who reasoned only with equations shows that students in the latter group were more confident about claiming, e.g., that the magnetic force should be perpendicular to the magnetic field. However, after performing the experiment, students who had reasonably good idea of the theory took the time to connect theory with non-intuitive experimental observation.  It is likely that introductory students need concrete experiences to make sense of these non-intuitive ideas and equations and connect them with physical phenomena.  Another finding is that reasoning involving more than one step, e.g., explanation for why the force between two long straight wires carrying current is in opposite directions is repulsive, was typically very challenging for a majority of students. Similarly, finding the net force when two separate forces, e.g., due to the electric and magnetic fields are present, was difficult for many students. Many students did not draw free body diagrams to reason systematically about each of the separate forces before finding the resultant force even though this approach is used by textbooks and instructors repeatedly in the context of linear and rotational dynamics in the first semester and in the context of electrostatic and magnetic forces in the second semester courses. Students often used gut feeling rather than systematic reasoning to answer such questions. Students should be explicitly taught the divide and conquer approach, i.e., how to solve a problem systematically by dividing it into sub-problems. Students should also be given opportunity to solve paired quantitative and conceptual problems which use the same principle of physics together so that they do not treat conceptual problems as guessing task. These findings and assessment tools can be used for developing curricular material that can improve students' understanding of these concepts.